\documentclass[12pt,preprint]{aastex}

\slugcomment{Submitted: ApJ, September 2008}

\shorttitle{PL Slope}
\shortauthors{Madore et al.}

\begin{document}

\medskip
\title{Concerning the Slope of the Cepheid Period-Luminosity Relation}
\medskip
\medskip
\medskip

\author{\bf Barry ~F. Madore \& Wendy L. Freedman} \affil{The Observatories \\ Carnegie Institution of
Washington \\ 813 Santa Barbara St., Pasadena, CA ~~91101\\}
\email{barry@ociw.edu, wendy@ociw.edu\\}




\begin{abstract}
We discuss the impact of possible differences in the slope of the
Cepheid Period-Luminosity relation on the determination of
extragalactic distances in the context of recent studies that suggest
changes in this slope. We show that the Wesenheit function $W = V - R
\times ((V-I)$, widely used for the determination of Cepheid
distances, is expected to be highly insensitive to changes in the
slope of the underlying (monochromatic) Period-Luminosity (PL)
relations. This occurs because the reddening trajectories in the
color-magnitude plane are closely parallel to lines of constant
period. As a result $W$-based Period-Luminosity relations have extremely
low residual dispersion, which is because differential (and total
line-of-sight) reddening is eliminated in the definition of $W$ and the
residual scatter due to a star's intrinsic color/position within the
Cepheid is also largely insensitive to $W$.  Basic equations are
presented and graphically illustrated, showing the insensitivity of $W$
to changes in the monochromatic $PL$ relations.

~

~

~

~

\end{abstract}

 
\section{INTRODUCTION}
There has recently been some concern raised over the possibility that
the slope of Cepheid Period-Luminosity ($PL$) relation is not
universal. Ngeow et al. (2005), Kanbur et al. (2007), Koen \& Siluyele
(2007) and other studies cited therein suggest that the optical $PL$
relation shows a break around ten days. These results are
controversial and are still being debated in the literature (see for
instance Benedict et al. 2008, and Pietrzynski et al. 2007). Given the
impact that Cepheids have traditionally had on the determination of
distances to galaxies it is important to confirm the effect, quantify
it, and then if it proves to be real, fully assess the impact of
it.\footnote{In that regard it does need to be said that for the extragalactic
distance scale the vast majority of studies have used Cepheids whose
periods are all longer than 10 days.} However, it is not the purpose of
this short contribution to attempt that assessment, rather we wish to
show explicitly that the reddening-free form of the Period-Luminosity
relation, the so-called Wesenheit function ($W$), can be demonstrated to
be largely immune to changes in the shape of the instability strip
that may then be reflected in changes in the slope of the $PL$
relation. Ngeow \& Kanbur (2005) have already demonstrated in one
instance (the Large Magellanic Cloud Cepheid sample) the robust nature
of W in this regard. In the following we show why this is more
generally true.

\section{The PLC}

We begin by noting that, for all stars, Stephan's Law in combination
with simple spherical geometry gives rise to a two-parameter mapping
of radius and effective temperature into total luminosity. By the
definition of effective temperature ($T_{eff}$) we have

$ L = 4\pi R^2 \sigma T^4_{eff} ~~~~~~~~~(1)$

Allowing a photometric color ((V-I)$_o$, say) to stand in for
temperature, and a monochromatic magnitude (M$_V$, say) to substitute
for bolometric luminosity, only one more observable is needed to
complete the transfer from theory to observation. For Cepheids, their
(easily observed) periods can stand in for their (hard to observe)
radii; where lines of constant radius in the color-magnitude plane
functionally substitute for lines of constant period.  The physically
motivated equation above then becomes the following equation based
exclusively on observables, while necessarily retaining the form of an
equation with two (and only two) independent variables: in this case
log(P) and $(V-I)_o$. Thus

$ M_V = \alpha $log(P)$ + \beta (V-I)_o + \gamma ~~~~~~~~~~~(2)$ 

It is important to note that both of these above equations describe
and cover the infinite plane of the color-magnitude diagram. There is
no period-luminosity relation in either of these equations. More
forcefully stated, Equation (2) is \underline{not} the
Period-Luminosity relation for Cepheids.  Any general correlation of
period and luminosity will only come {\it through a constraint} of
these equations. That constraint, in the context of Cepheids, is
imposed by the existence of an instability strip within which the
natural period of oscillation is excited, and manifest as a periodic
change of luminosity. The instability strip, expressed as an
additional and external mathematical constraint on the above
equations, produces the observed period-luminosity relations by
narrowing down the range of luminosities seen at any given period. The
slope of the red and blue boundaries to the instability strip as they
cut lines of constant period determines the slope of the PL
relation. This was first illustrated and emphasized in Madore \&
Freedman (1991), their Figure 3.

Before advancing further, we introduce the Wesenheit function $W = V
- R_{VI} \times (V-I)$.  This specific combination of magnitude and
color is explicitly constructed so as to
mathematically guarantee that their sum will be numerically
independent of the total amount of interstellar extinction. With
foreknowledge that (by definition) $A_V = R_{VI} \times E(V-I)$, and
again having the added definitions that $V = V_o + A_V$ and $(V-I) =
(V-I)_o + E(V-I)$, with simple algebra

$W = V - R_{VI} \times (V-I) $

\noindent
then

$= V_o + A_V - R_{VI} \times (V-I)_o - R_{VI} \times E(V-I)$

\noindent
which upon regrouping terms becomes

$ W = [V_o  - R_{VI} \times (V-I)_o] + [A_V - R_{VI} \times E(V-I)]$

\noindent
By construction and definition the last two terms sum to zero, leaving 

$ W = V_o  - R_{VI} \times (V-I)_o = W_o$

In other words, $W$, as constructed from observed (reddened) colors and
(extincted) magnitudes, is numerically equivalent to $W_o$, as it would
be calculated from intrinsic colors and intrinsic magnitudes. This
occurs, it must be noted, without any explicit knowledge of the
reddenings and/or extinctions involved. Simply put, $W$ is
reddening-independent.

The geometrical interpretation of $W$ is illuminating. As Figure 1
illustrates, lines of constant $W$ fall, by construction, identically
along reddening trajectories having a slope of $R_{VI} =
A_V/E(V-I)$. Whether a star is at position B because it was reddened
from its intrinsic position at A, or whether it is there because that
is its intrinsic color and magnitude, is irrelevant: $W$ collapses the
color-magnitude diagram in such a way that the distinction usually
made by reddening is moot. All stars along a given reddening
trajectory will have the same value of $W$. Stars with a given value
of $W$ may have any value of reddening.

We now interface $W$ to Equation 2. First of all

$ M_W = M_V  - R_{VI} \times (V-I)_o$

\noindent
and so

$ M_W = \alpha $log(P)$ + (\beta - R_{VI}) \times (V-I)_o + \gamma$ 

Clearly if $(\beta - R_{VI}) = 0.0$ then $W$ will be not only
reddening-free but it will also be a dispersionless function of period
alone, with slope = $\alpha$. Indeed, it appears observationally that
$\beta$ and $R_{VI}$ must be quite similar in magnitude given the fact
that the dispersion in $W$ is found to be very small (i.e., on the
order of $\pm$0.08~mag).

\subsection{PERIOD-LUMINOSITY RELATIONS}

\subsection {The Mathematical Representation}

Let us take the mean ridge-line relation between magnitude and color
(defining the instability strip that is to act as the constraint on
the $PLC$) to be represented by $M_V(mean) = S \times (V-I)(mean)$. It
then follows from Equation 2 that 

$M_V(mean) = \alpha ~[S/(S - \beta)] ~$log(P)$ - \gamma ~[S/(S - \beta)]$ 

\noindent
and 

$W(mean) = \alpha  ~[(S - R_{VI})/(S - \beta)] ~$log(P)$ - \gamma ~[(S - R_{VI})/(S - \beta)]$. 

\noindent
Scatter in these mean (period-luminosity) relations will be determined
by the color width of the instability strip multiplied by (the
projection factor) $\beta$, in the case of monochromatic $PL$
relations, and multipled by ($\beta - R$) in the case of the Wesenheit
functions. As noted above, if $\beta$ and $R$ are of similar magnitude
then the scatter in the Wesenheit function will be less than the
scatter in the corresponding $PL$ relations by the factor $(\beta -
R)/\beta$. Should $\beta$ equal $R$ then $W$ would be dispersionless.

If the instability strip (the constraint imposed upon Equation 2)
moves in a way that changes the slope of the mean period-luminosity
relation through a change in $\Delta S$ this will result in a change
in the slope of the mean W-logP relation, $\Delta A_W$ =  $\alpha
(R - \beta)/(S - \beta)^2] \times \Delta S$. For the same tilt in
the instability strip $\Delta S$, the slope in the monochromatic PL
relation will change by $\Delta A = -\alpha \beta/(S - \beta)^2 \times
\Delta S$. This means that a 10\% change (say) in the slope of the
V-band $PL$ relation mapped to $W$ would be reduced by a factor of
$-(R - \beta)/\beta$. This is the same factor that results in the
decreased scatter observed in $W$ as compared to the scatter at fixed
period in the V-band $PL$ relation. In the extreme case where $\beta$
equals $R$ then $W$ would not change its slope at all for any change
in tilt of the instability strip.

\subsection{The Graphical Representation}

The central portion of Figure 1 shows a V-(V-I) color-magnitude
diagram with two possible realizations of the Cepheid instability
strip cutting across lines of constant period as laid down by the
underlying period-luminosity-color relation. The strip to the left is
significantly more vertical that the strip to the right. Both
instability strips share the same zero point in this realization, but
they could differ without changing the conclusions of this
paper. Fiducial points are marked in both strips by open and filled
circles. The filled circles in the more highly inclined strip mark the
blue and red borders of the instability strip at two distinct
periods. The open circles in the left-hand strip mark the same color
boundaries at precisely the same two periods. The downward-sloping
dashed lines crossing the entire figure are lines of constant
period. 

The dashed horizontal lines carry the individual data points
in the tilted instability strip over to the right and into its
corresponding $PL$ relation. Data from the steep instability strip map
to the left $PL$ relation. Clearly by tilting the instability strip one
generates $PL$ relations with different slopes. However, by projecting
the instability strips along lines of constant period, as is done by
following the broken lines to the upper left panel, one finds a
dispersionless $PL$ relation as defined by $W_{\beta}$ which, in addition
to having no scatter, is incapable of distinguishing between stars
originating either in the vertical or the tilted instability strip. 

The solid lines emanating from the filled data points in the tilted
instability strip show the trajectories of reddening lines. Their
slight divergence from lines of constant period results in a composite
W-logP relation that has some residual scatter and a slight dependence
of its slope on the originating slant of the instability strip. Unlike
the gross differences seen in the (left and right) monochromatic PL
relations the divergence in the $W_{\beta}$ relations can hardly be
resolved and require an exploded view as given in Figure 2 to see the
impact.

\subsection{The Numerical Representation}

It has been claimed (Sandage \& Tammann 2008, their Table 1) that
variations in the slope of the V-band $PL$ relation are found at the 5\%
level around a median value of about -2.75 among nearby galaxies
(excluding the poorly populated Sextans A/B data point) in which
Cepheids have been discovered. If one were to rely exclusively on
V-band $PL$ relations for Cepheid distances that might be cause for
concern. But such is not the case. Virtually all modern studies of the
distances to galaxies as gauged by optical observations of Cepheids
use $W$.

For the sake of the following calculation we assume that $R_{VI}$ =
2.45 (as in Freedman et al. 2001) and that $\beta = 3.0$ (as derived,
for example, from the sample of Galactic Cepheids observed by Benedict
et al. 2008, illustrated in Figure 3).  In this case the reduction
factor $V$ to $W$ in the slope change as given above is $-(R -
\beta)/\beta = -(3.0 - 2.45)/2.45 = -0.3$. That is, a 5\% change in
slope seen in V-band $PL$ relation translates into a 1.5\% change in
slope of the mean $W$-logP relation. Had $\beta$ and R been
numerically identical then there would have been absolutely no impact
of a change in the disposition of the instability strip on the slope
of the mean $W$-logP relation. And indeed this would be almost exactly
the case if we had adopted the value of $\beta$ = 2.43 as advocated by
Sandage \& Tammann (2008) based on a study of LMC Cepheids. In that
case the 5\% slope changes in the optical would reduce to a less that
two-tenths of a percent change in the slope in $W$.  We do not believe
that either $\beta$ or $R$ are currently known to 10\% precision
themselves, so the exercise must be taken cautiously, being
illustrative rather than definitive.  The point stands however that
the slope of $W$ is clearly going to be relatively impervious to
possible changes in the slopes at optical wavelengths.

\section{CONCLUSIONS}

Not only does the scatter intrinsic to the $W$-logP relation collapse
with respect to the optical $PL$ relations, but the slope of the
$W$-logP relation is also very insensitive to changes in slope of the
originating optical $PL$ relations. Changes in the tilt of the
instability strip that naturally explain and will give rise to changes
in slopes of the monochromatic $PL$ relations are greatly diminished
in their overall impact in the projection of data into the $W$-logP
plane. The Wesenheit function is defined to be reddening-free; however
its slope is additionally and coincidently very resistant to changes
in the slopes of the originating monochromatic, optical $PL$
relations.  The demonstration by Ngeow \& Kanbur (2005) that the
Wesenheit function for LMC Cepheids does not show any evidence for a
break\footnote{The referee points out that Koen, Kanbur \& Ngeow
(2007) also find a linear Wesenheit/PLC relation in the LMC, but note
that with more data non-linearities could be measureable} in slope
over the full range of observed periods is a predictable and universal
feature of $W$. It naturally follows from the closely coincident
slopes of the lines of constant period and reddening trajectories.

Since $W$ is already the default method for determining true moduli to
nearby galaxies when using Cepheids observed at optical wavelengths,
we conclude that slope changes in the $V$-band $PL$ relations,
confirmed or not, will have minimal impact on the Wesenheit function,
or the distance scale based upon it. This point has already been made
by Ngeow \& Kanbur (2006) where they suggest that slope changes may
affect determinations of $H_o$ at the 1-2\% level.

\acknowledgements This paper was written while the authors were in
residence at L'Institute d'Astrophysique and L'Observatoire de
Paris. We each thank Professor Daniel Egret, Director of the Paris
Observatory, and Dr. Stephan Charlot at IAP for the opportunity to be
with them for an extended visit and for their kind hospitality and
support during that time. Finally, we thank the referee for carefully
reading the manuscript and alerting us to transcription errors in
the first draft of this paper.

\vfill\eject
\noindent
\centerline{\bf References \rm}
\vskip 0.1cm

\par\noindent 
Benedict, G. F., et al. 2008, \aj, 133, 1810

\par\noindent 
Kanbur, S.M., Ngeow, C., Nanthakumar, A., \& Stevens, R.  2007, \pasp, 119, 512

\par\noindent 
Koen, C., Kanbur, S.M., \& Ngeow, C.  2007, \mnras, 380, 1440

\par\noindent 
Madore, B.F. \& Freedman \, W.L. 1991, \pasp, 103, 933

\par\noindent 
Freedman, W.L., \& Madore, B.F. 2009 (in preparation)

\par\noindent 
Ngeow, C., \& Kanbur, S.M. 2005, \mnras, 360, 1033 

\par\noindent 
Ngeow, C., \& Kanbur, S.M. 2006, \apj, 650, 180

\par\noindent 
Ngeow, C., Kanbur, S.M., \& Nanthakumar, A. 2008, \aa, 477, 621

\par\noindent 
Pietrzynski, G. et al. 2007, \aj, 134, 594

\par\noindent 
Sandage A.R., \& Tammann, G.A. 2008 (astro-ph 0803.3836)

\vfill\eject
\vskip 0.75cm

\begin{figure}
\includegraphics [width=13cm, angle=270] {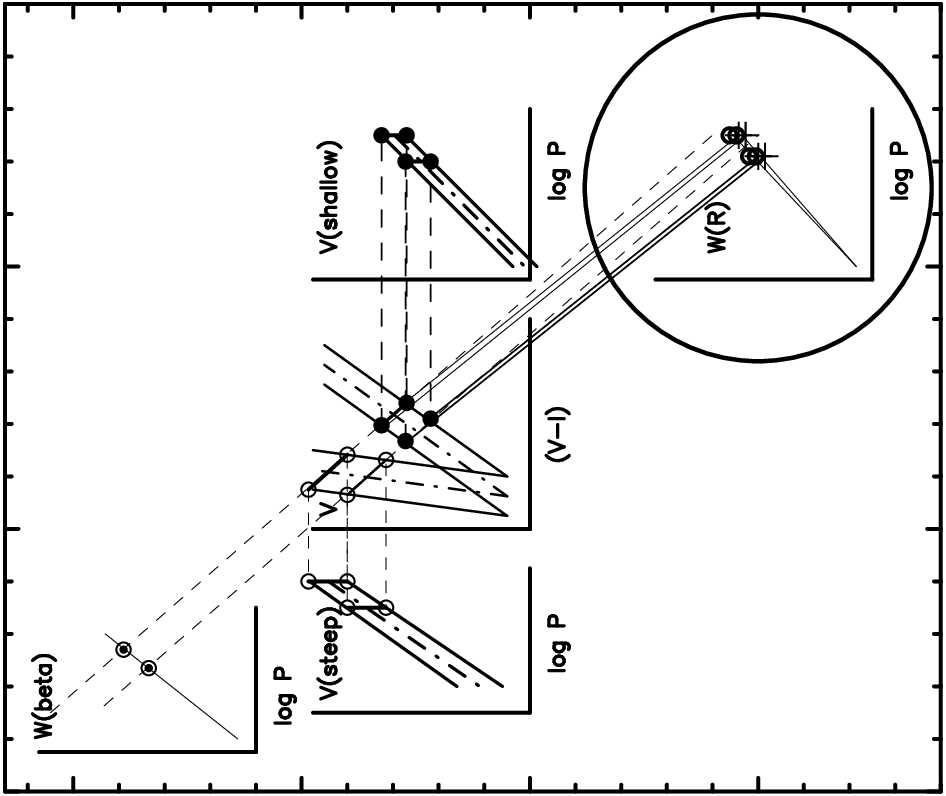} \vfill
\medskip
\medskip
\medskip

\par\noindent \bf Fig. 1 \rm -- The geometrical mapping of the Cepheid
instability strip into period-luminosity relations, and the
consequences of tipping the instability strip.

\end{figure}

\begin{figure}
\includegraphics [width=10cm, angle=270] {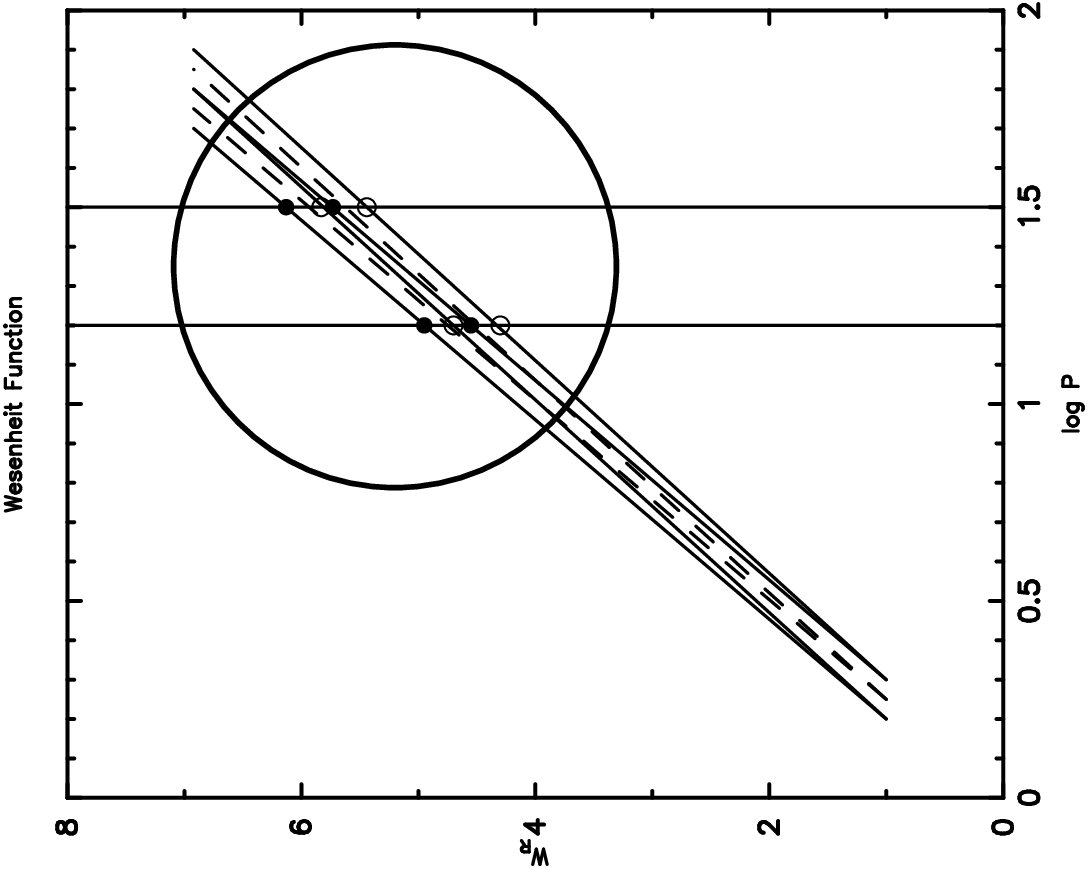} \vfill
\medskip
\medskip
\medskip

\par\noindent \bf Fig.  2 \rm -- An expanded view of the $W_{R}$-logP
relation as generated by two instability strips of different tilt. The
slight offset in slopes between the lines of constant period and
reddening lines leads to dispersion in W and a (very) slight relative
tilt of the two relations. Both the dispersion and change in slope
collapse to zero linearly with the difference ($\beta$ - $R_V$).
\end{figure}

\begin{figure}
\includegraphics [width=12cm, angle=270] {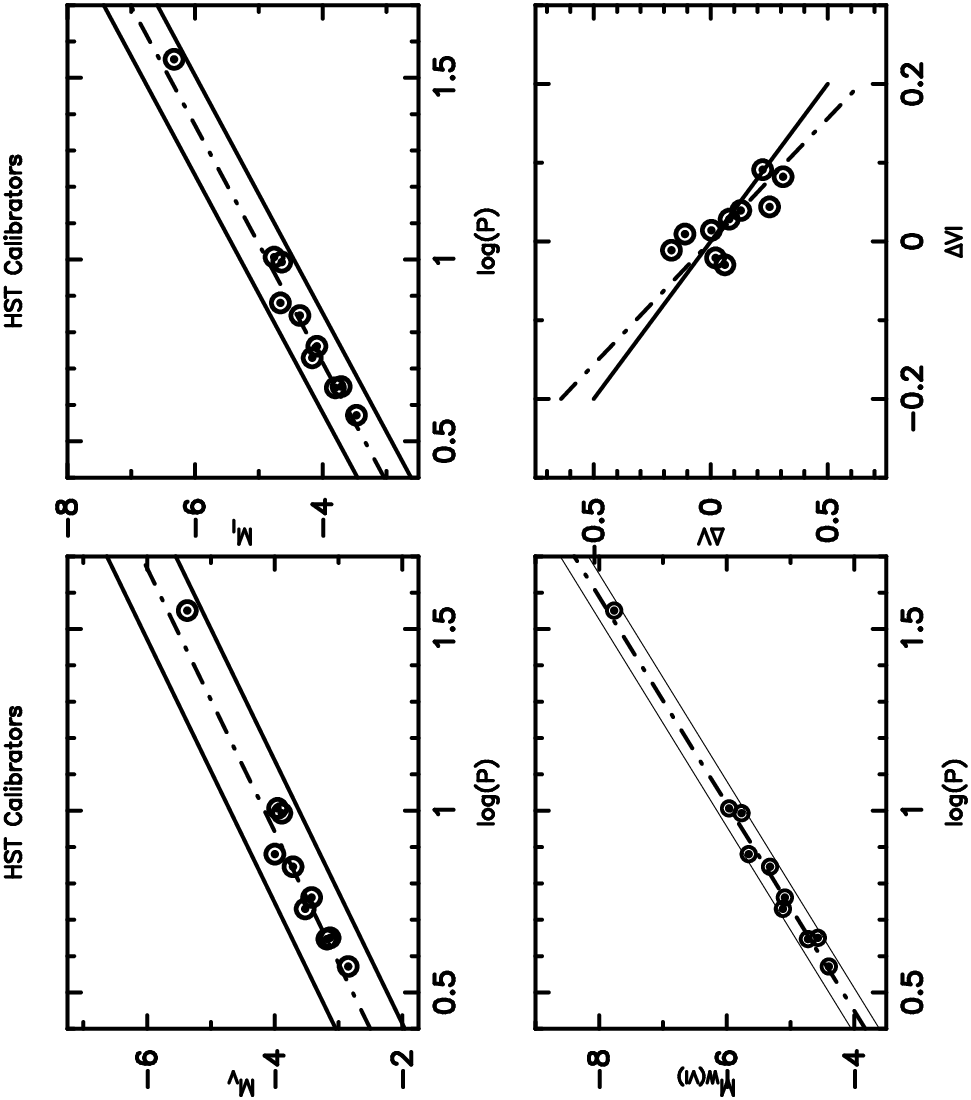} \vfill
\medskip
\medskip
\medskip

\par\noindent \bf Fig.  3 \rm -- Galactic Cepheid calibrators from
Benedict et al. (2008). Upper panels: V and I-band Period-Luminosity
relations. Lower left panel: W(VI) $PL$ relation. Lower right panel:
the correlation of V and (V-I) residuals from their respective $PL$ and
PC relations. The solid line is a the trajectory expected for differential
reddening; the dashed line is the best-fit regression. The difference
between the two is not statistically significant; see Freedman \&
Madore (2009) for details.

\end{figure}

\end{document}